\documentclass[aps,prd,superscriptaddress,11pt,onecolumn]{revtex4}
\usepackage{amssymb}
\usepackage{amsmath}
\usepackage{amsfonts}
\usepackage{graphicx, float}
\usepackage{graphicx, epsfig}
\usepackage{color}
\usepackage{enumerate}

\newcommand{\bq}{\begin{equation}}
 \newcommand{\eq}{\end{equation}}
 \newcommand{\bqn}{\begin{eqnarray}}
 \newcommand{\eqn}{\end{eqnarray}}
 \newcommand{\nb}{\nonumber}
 \newcommand{\lb}{\label}
\newcommand{\beq}{\begin{equation}}
\newcommand{\eeq}{\end{equation}}
\newcommand{\bea}{\begin{eqnarray}}
\newcommand{\eea}{\end{eqnarray}}

\begin{document}

\title{Resolving the information loss paradox from the five-dimensional minimal supergravity black hole}

\author{Behnam Pourhassan}\email{b.pourhassan@candqrc.ca}
\affiliation{School of Physics, Damghan University, Damghan, 3671641167, Iran.}
\affiliation{Canadian Quantum Research Center, 204-300232 AveVernon, BCV1T2L7, Canada.}

\begin{abstract}

{\bf Abstract:} In this paper, we study the Hawking tunneling radiation of the charged rotating black hole in five-dimensional minimal supergravity theory by using the semiclassical Hamilton-Jacobi equation. By using two separated ways we obtain the corrected entropy of the black hole. Equality of results gives us a special condition that may solve the information loss paradox. Then, we focus on the phase transitions, and the results show that if the effects of thermal fluctuations are incorporated in the entropy, the black hole is unstable, while there are phase transitions according to the sign-changing behavior of the black hole specific heat. We find that, in presence of thermal fluctuations, the black holes of five-dimensional minimal supergravity behave like the black holes in Horava-Lifshitz gravity hence the second-order phase transition is possible.

\end{abstract}

\maketitle
\newpage
\section{Introduction}

Inspired by several developments in higher dimensional string theory, Myers and Perry (MP) proposed a rotating black hole of dimensions $d=2n+1>4,$ with the aim of better understanding the string theory itself \cite{mp,mp1}. For $n=2$, the metric possesses three Killing vectors, corresponding to translations in coordinates $t$,  $\phi$ and $\psi$. The MP space-time possesses the gravitational structures of singularity and the event horizon. In that special case no violation of the weak cosmic censorship conjecture occurs around the MP black hole \cite{111}. Numerous physical features of MP black holes have been investigated already including the stability issues in \cite{222}, as particle accelerators \cite{333} and as sources of Hawking radiations \cite{mp2,mp3}. Moreover, statistics of MP black holes were investigated in \cite{mp4} and its relation with the Kerr-G\"{o}del black hole  had been found \cite{G1,G2,G3,G4}.
The black hole entropy gets necessary correction due to thermal fluctuations of space-time resulting in the logarithmic correction term as a leading order to the entropy \cite{2,3}. In that case, thermodynamics of higher dimensional black holes including MP black hole with higher order thermal fluctuations has been studied in \cite{Rang, MPlog}. We would like to study corrected thermodynamics, Hawking radiation and stability of the 5D minimal supergravity black hole, which is a Myers-Perry black hole containing the electric charge \cite{mp, cho}.\\
One of the outstanding contributions of Stephen Hawking is the discovery of evaporation of black holes via emission of thermal radiation \cite{haw1,haw2}. The black hole is a hot object with temperature proportional to its surface gravity. This radiation is composed of all sorts of particles such as bosons and fermions. More recently it has been argued that this radiation is approximately thermal and hence not exactly Planckian \cite{vis}. The phenomenon of Hawking radiation has been obtained via various approaches of quantum field theories in curved backgrounds \cite{Muk} and hence must be an integral part of any candidate theory of quantum gravity such as string theory \cite{string}. In literature, the mechanism of Hawking radiation is proposed as a process of quantum tunneling at the horizon, by which the horizon temperature and tunneling probability can be calculated. In this scheme, the Klein-Gordon equation governing the dynamics of a scalar particle (spin-0) is cast as Hamilton-Jacobi scheme equation \cite{pady}. The Dirac fermion tunneling is proposed by Kerner and Mann \cite{mann}, but their method considers the spin-up and spin-down cases respectively. By following their work, we can get Hamilton-Jacobi equation from Dirac equation directly, so that the Hamilton-Jacobi method also can be applied in Dirac case \cite{lin1,lin2,LK}. It is easy to show that the Hamilton-Jacobi equation could be derived by all the kinds of field equation, including spin-$0$ (Klein-Gorden field), spin-$1/2$ (Dirac field), spin-$1$ (Maxwell and Proca fields), spin-$2$ (for example gravitational wave), spin-$(n-1/2)$ (Rarita-Schwinger field) and spin-$n$ case (higher spin boson field). Therefore, the Hawking tunneling radiation could be investigated by Hamilton-Jacobi equation uniformly.\\
One of our aims is to deduce the tunneling rate and temperature in the 5D minimal supergravity black hole. We shall study the evolution of temperature with the size of the black hole. On the other hand, a non-extremal rotating black hole with the electric charge in minimally gauged five dimensional supergravity already proposed \cite{cho}. In that case the thermodynamic stability analysis has been done for the charged rotating black holes \cite{1512.04181}.
In this work, we shall consider the thermal fluctuations corrections of black hole entropy to investigate the phenomenon of phase transitions. It would be carried out by checking the behavior of pressure against volume of the black hole, while also analyzing the sign-changing behavior of specific heat of the black hole. Already, the quantum space-time fluctuation corrections to the Bekenstein-Hawking area law
for black hole entropy was derived which leads to the logarithmic corrections of a universal nature \cite{0002040}.  Also, $P-V$ criticality of some kinds of black hole has been investigated under effects of thermal fluctuations, for example see Refs. \cite{86, 111111}.\\
This paper is organized as follows, in the next section we review some aspects of 5D minimal supergravity black hole which will be useful for another sections. Here, we give numerical analysis of horizon structure. Then, in section III Hawking tunneling radiation of the black hole is investigated and black hole temperature studied numerically. In section IV, we employ the logarithmic correction coming from thermal fluctuations and use the log corrected entropy to investigate phase transitions. In that case, we compare corrected entropies of black hole and particles to obtain condition where information loss paradox solved. Finally in section V we give conclusion and summary of results.

\section{Five dimensional minimal supergravity black hole}
\begin{figure}[h!]
 \begin{center}$
 \begin{array}{cccc}
\includegraphics[width=75 mm]{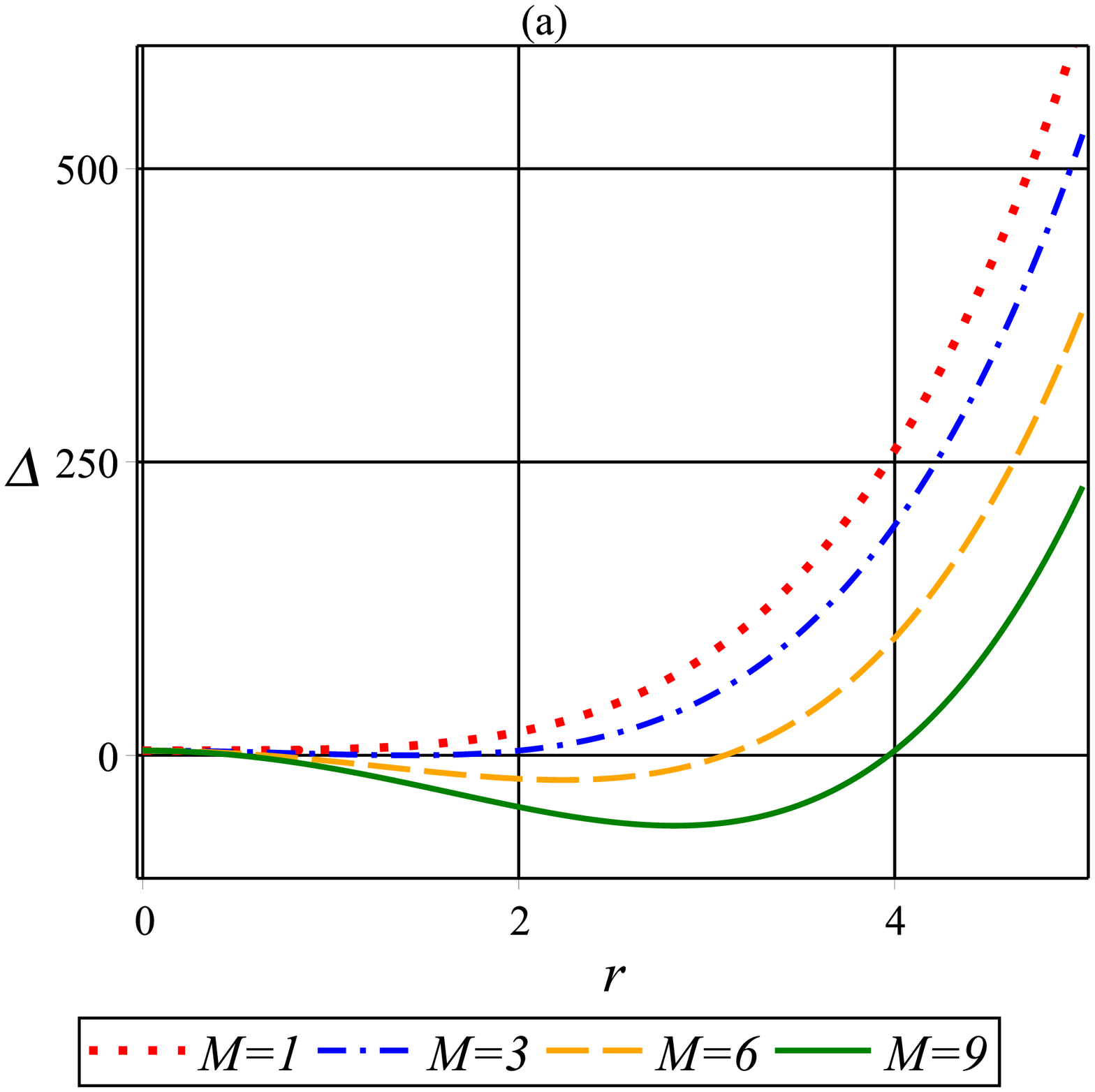}\includegraphics[width=75 mm]{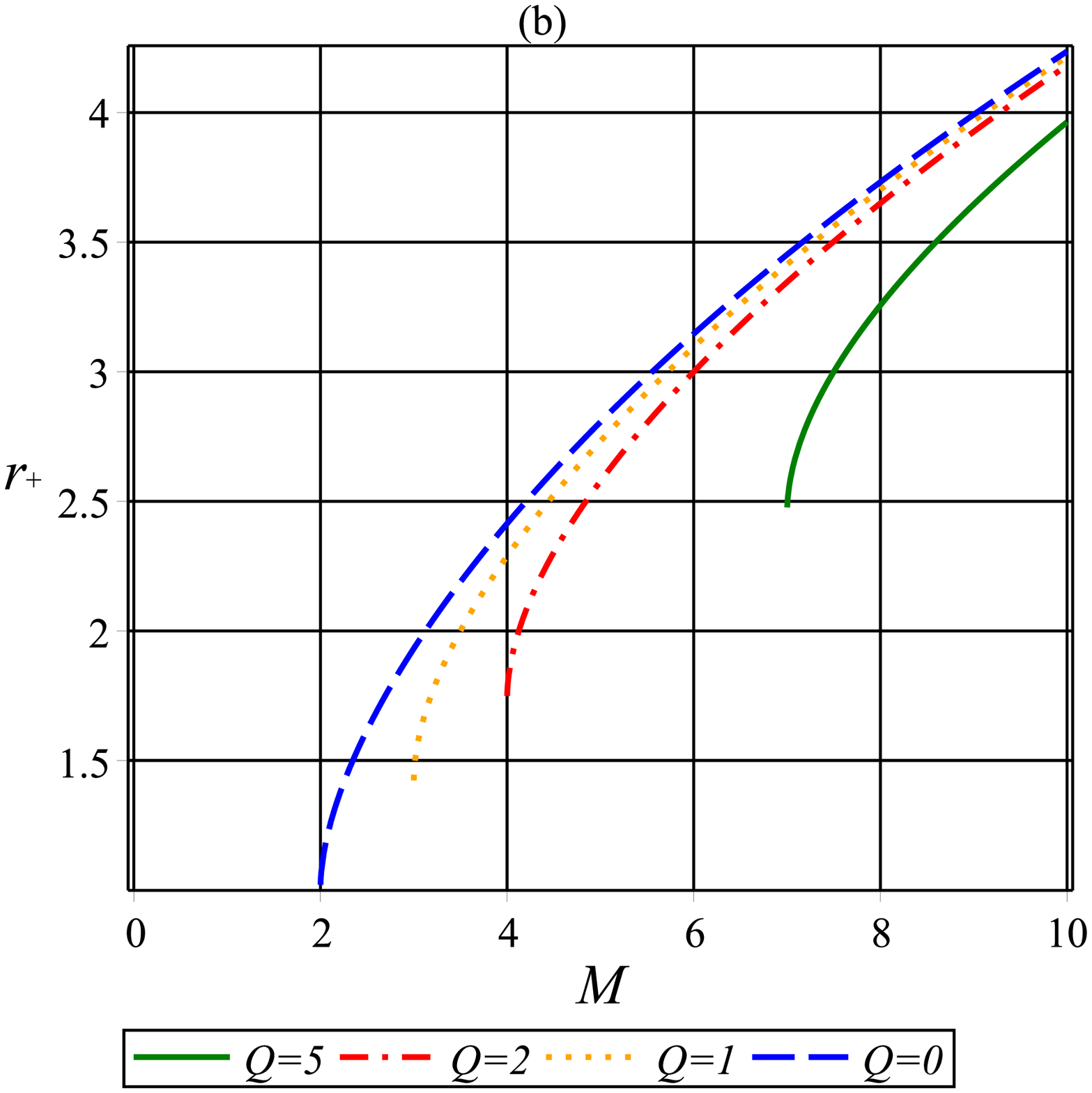}
 \end{array}$
 \end{center}
\caption{(a) $\Delta$ in terms of $r$ for $Q=a=b=1$. $r_{+}$ in terms of $M$ for $a=b=1$.}
 \label{1}
\end{figure}

The action of five-dimensional minimal supergravity (the bosonic sector of
supergravity) is given by \cite{mp,jcap},
\begin{equation}\label{0}
I=\int{d^{5}x\sqrt{-g}\left(R-\frac{1}{4}F_{\alpha\beta}F^{\alpha\beta}
+\frac{1}{12\sqrt{3}}\epsilon^{\mu\nu\alpha\beta\gamma}F_{\mu\nu}F_{\alpha\beta}\mathbf{A}_{\lambda}\right)}.
\end{equation}
The bosonic equations of motion may be written as \cite{B1,B2},
\begin{eqnarray}
R_{\alpha\beta}+2F_{\alpha\gamma}F_{\beta}^{\gamma}-\frac{1}{3}g_{\alpha\beta}F^{2}&=&0\nonumber\\
dF+\frac{2}{\sqrt{3}}\epsilon^{\mu\nu\alpha\beta}F_{\mu\nu}F_{\alpha\beta}&=&0.
\end{eqnarray}
The general solution to above field equations describing a charged rotating black hole with two rotation parameters $a$ and $b$ and electric charge $Q$ with mass parameter $M$ is given by \cite{mp,cho,jcap},
\begin{eqnarray}
ds^2&=&-(dt-a\sin^2\theta d\phi-b\cos^2\theta d\psi)[f(dt-a\sin^2\theta d\phi-b\cos^2\theta d\psi)\nonumber\\&&
+\frac{2Q}{\Sigma}(b\sin^2\theta d\phi+a\cos^2\theta d\psi)]+\Sigma(\frac{r^2dr^2}{\Delta}+d\theta^2)\nonumber\\&& +\frac{\sin^2\theta}{\Sigma}[adt-(r^2+a^2)d\phi]^2+\frac{\cos^2\theta}{\Sigma}[bdt-(r^2+b^2)d\psi]^2\nonumber\\&&+\frac{1}{r^2\Sigma}[abdt-b(r^2+a^2)\sin^2\theta d\phi-a(r^2+b^2)\cos^2\theta d\psi]^2,
\end{eqnarray}
where the metric components are specified by
\begin{eqnarray}
f&=&\frac{(r^2+a^2)(r^2+b^2)}{r^2\Sigma}-\frac{2M\Sigma-Q^2}{\Sigma^2},\nonumber\\
\Sigma&=&r^2+a^2\cos^2\theta+b^2\sin^2\theta,\nonumber\\
\Delta&=&(r^2+a^2)(r^2+b^2)+2abQ+Q^2-2Mr^2.
\end{eqnarray}
The horizons of the above metric follow from the relation $\Delta=0$, i.e.
\begin{equation}\label{3}
r_{\pm}^2=\frac{1}{2}[-(a^2+b^2-2M)\pm\sqrt{(a^2+b^2-2M)^2-4(ab+Q)^2}],
\end{equation}
with $r_{+}\geq r_{-}$. It is clear that there are two horizons if
\begin{equation}\label{Mass}
M>\frac{{a}^{2}}{2}+ab+\frac{{b}^{2}}{2}+Q.
\end{equation}
In the case of equality, we define $M_{c}\equiv \frac{{a}^{2}}{2}+ab+\frac{{b}^{2}}{2}+Q$, the black hole is extremal where $r_{+}=r_{-}=\sqrt{ab+Q}$. Otherwise, there is a naked singularity. In Fig. \ref{1} (a) we can see horizon structure of 5D minimal supergravity black hole. We can see that two horizons are possible if $M>M_{c}$. Blue dash-dotted line represent extremal case where $M=M_{c}$, while red dotted line represents only singularity at the origin ($M<M_{c}$). It means that there is a lower bound for the black hole mass ($M_{c}$) which corresponds to the lower bound for the event horizon radius ($r_{+}$). Also, in Fig. \ref{1} (b) the event horizon is an increasing function of the black hole mass as expected. The lower bounds of $M$ and $r_{+}$ are shown for some values of the black hole charge.

The angular speeds of the stationary observers at the event horizons are :
\begin{equation}
\Omega_{a+}=\frac{2\pi^2}{A}\frac{a(r_+^2+b^2)+bQ}{r_+},~~~ \Omega_{b+}=\frac{2\pi^2}{A}\frac{b(r_+^2+a^2)+aQ}{r_+},
\end{equation}
where $A$ denotes the area of the event horizon given by
\begin{equation}
A=\frac{2\pi^2[(r_+^2+a^2)(r_+^2+b^2)+abQ]}{r_+}.
\end{equation}
%Moreover, the temperature is specified by
%\begin{equation}\label{T7}
%T=\frac{\kappa}{2\pi}=\frac{\pi(r_+^2-r_-^2)}{A}
%\end{equation}
The electrostatic potential  of the horizon relative to an observer lying at infinity is
\begin{equation}
\Phi_H=\frac{\sqrt{3}\pi^2 Qr_+}{A}.
\end{equation}

\section{Hawking Tunneling Radiation}

The classical black hole cannot radiate particles because any particle inside the horizon fails to escape from black hole. However, Hawking considered the quantum effect near the horizon of black hole and found that black hole can emit radiation, which is named as Hawking radiation \cite{haw1,haw2}. This important work developed the connections between gravity, quantum theory and thermodynamics, and now offers a pathway towards the theory of quantum gravity.
Parikh and Wilczek studied Hawking radiation by tunneling process \cite{Tunneling1,Tunneling2,Tunneling3}, later Hamilton-Jacobi tunneling method was also proposed \cite{HJ1,HJ2,lin3}. In this section we will study the tunneling radiation of bosonic and fermionic particles of diverse spins following \cite{lin4}. It is easy to show that the field equations for fermions and bosons can be written as Hamilton-Jacobi equation in semi-classical approximation for fermions as well as bosons \cite{LK,lin1,lin2,lin3,lin4}. The key assumption of this method is that we can neglect any change of black hole angular momentum due to the emitted particles spin \cite{m1,m2}. It is a good approximation for the black holes with mass much larger than the Planck mass and vanishing angular momentum. Also, from the statistical mechanics point of view, particles with opposite spin will be emitted in the equal numbers, so angular momentum of the black hole keep constant.\\
Above discussions show that various kinds of quantum fields equation can get Hamilton-Jacobi equation by semi-classical dynamical approximation, so that the Hawking tuneling radiation with all kind of spins can be described by Hamilton-Jacobi equations. Consequently, we will use Hamilton-Jacobi equation to calculate Hawking radiation of 5D minimal supergravity black hole.\\
On the other hand, the black hole's inversion metric is given by,
\begin{eqnarray}\label{InvM}
g^{00}&=&\frac{1}{\Sigma}\left[a^2\sin^2\theta+b^2\cos^2\theta-\frac{{\cal A}^2}{r^2\Delta}+\frac{a^2b^2}{r^2}\right]\nonumber\\
g^{33}&=&\frac{1}{\Sigma}\left[\frac{1}{\sin^2\theta}-\frac{{\cal B}^2}{r^2\Delta}+\frac{b^2}{r^2}\right]\nonumber\\
g^{44}&=&\frac{1}{\Sigma}\left[\frac{1}{\cos^2\theta}-\frac{{\cal C}^2}{r^2\Delta}+\frac{a^2}{r^2}\right]\nonumber\\
g^{03}&=&\frac{1}{\Sigma}\left[a-\frac{{\cal AB}}{r^2\Delta}+\frac{ab^2}{r^2}\right]\nonumber\\
g^{04}&=&\frac{1}{\Sigma}\left[b-\frac{{\cal AC}}{r^2\Delta}+\frac{ba^2}{r^2}\right]\nonumber\\
g^{34}&=&\frac{1}{\Sigma}\left[-\frac{{\cal BC}}{r^2\Delta}+\frac{ab}{r^2}\right]\nonumber\\
g^{11}&=&\frac{\Delta}{\Sigma r^2}\nonumber\\
g^{22}&=&\frac{1}{\Sigma}
\end{eqnarray}
where
\begin{eqnarray}\label{InvMa}
{\cal A}&=&a^2b^2+abQ+(a^2+b^2)r^2+r^4\nonumber\\
{\cal B}&=&bQ+ab^2+ar^2\nonumber\\
{\cal C}&=&aQ+ba^2+br^2,
\end{eqnarray}
and electromagnetic potential
\bqn
\lb{InvMb}
\mathbf{A}=-\frac{\sqrt{3}Q}{2\Sigma}\left(dt-a\sin^2\theta d\phi-b\cos^2\theta d\psi\right)
\eqn
Let us separate the action of Hamilton-Jacobi equation $\mathcal{S}=-\omega t+R(r)+j_a\phi+j_b\psi+Y(\theta)$, where $\omega$ is the frequency of field particle, while $j_a$ and $j_b$ are magnetic quantum numbers. Therefore, the Hamilton-Jacobi equation becomes
\bqn
\lb{InvMc}
&&\frac{\Delta}{r^2}\left(\frac{dR(r)}{dr}\right)^2-\frac{\left({\cal A}\omega+{\cal B}j_a+{\cal C}j_b-\frac{\sqrt{3}}{2}qQr^2\right)^2}{r^2\Delta}+\frac{\left(ab\omega+bj_a+aj_b\right)^2}{r^2}+m^2r^2\nb\\
&&+\left(\frac{dY(\theta)}{d\theta}\right)^2+\left(a\sin\theta+\frac{1}{\sin\theta}\right)^2+\left(b\cos\theta+\frac{1}{\cos\theta}\right)^2+m^2a^2\cos^2\theta+m^2b^2\sin^2\theta=0
\eqn
so that the radial equation becomes
\bqn
\lb{InvMHJa}
\frac{\Delta}{r^2}\left(\frac{dR(r)}{dr}\right)^2-\frac{\left({\cal A}\omega+{\cal B}j_a+{\cal C}j_b-\frac{\sqrt{3}}{2}qQr^2\right)^2}{r^2\Delta}+\frac{\left(ab\omega+bj_a+aj_b\right)^2}{r^2}+m^2r^2+C_L=0
\eqn
where $C_L$ is a constant, and than we get
\bqn
\lb{InvMHJb}
R_\pm(r)=\pm\int \frac{dr}{\Delta}\sqrt{\left({\cal A}\omega+{\cal B}j_a+{\cal C}j_b-\frac{\sqrt{3}}{2}qQr^2\right)^2-\Delta\left(ab\omega+bj_a+aj_b\right)^2-\Delta\left(r^4m^2+r^2C_L\right)}.
\eqn
Near horizon $r_+$, we have $\Delta\sim\Delta'(r_+)(r-r_+)$. According to Residue theorem, we integrate $r$ from inside of horizon outside in complex region, and above integration is simplified as
\begin{eqnarray}\label{InvMHJc}
R_\pm(r)&=&\pm\left. \frac{1}{\Delta'(r_+)}\left({\cal A}\omega+{\cal B}j_a+{\cal C}j_b-\frac{\sqrt{3}}{2}qQr^2\right)\right|_{r=r_+}\int\frac{dr}{r-r_+}\nonumber\\
&=&\pm \left. \frac{i\pi}{\Delta'(r_+)}\left({\cal A}\omega+{\cal B}j_a+{\cal C}j_b-\frac{\sqrt{3}}{2}qQr^2\right)\right|_{r=r_+}.
\end{eqnarray}
Therefore, the Hawking radiation tunneling rate is given by
\bqn
\lb{TunnelingA}
\Gamma=e^{-2[\Im(R_+)-\Im(R_-)]}=\exp\left[-\frac{\omega-\omega_0}{T}\right],
\eqn
where Hawking temperature and the chemical potential respectively are given by
\begin{eqnarray}\label{TunnelingB}
T&=&\left.\frac{\Delta'}{4\pi{\cal A}}\right|_{r=r_+}=\frac{r_+\left(2r_+^2+a^2+b^2-2M\right)}{2\pi\left(a^2b^2+abQ+(a^2+b^2)r_+^2+r_+^4\right)}\nonumber\\
\omega_0&=&\left.\frac{\frac{\sqrt{3}}{2}qQr_+^2-{\cal B}j_a-{\cal C}j_b}{\cal A}\right|_{r=r_+}\nonumber\\
&=&\frac{\frac{\sqrt{3}}{2}qQr_+^2-(bQ+ab^2+ar_+^2)j_a-(aQ+ba^2+br_+^2)j_b}{a^2b^2+abQ+(a^2+b^2)r_+^2+r_+^4}.
\end{eqnarray}

\begin{figure}[h!]
 \begin{center}$
 \begin{array}{cccc}
\includegraphics[width=70 mm]{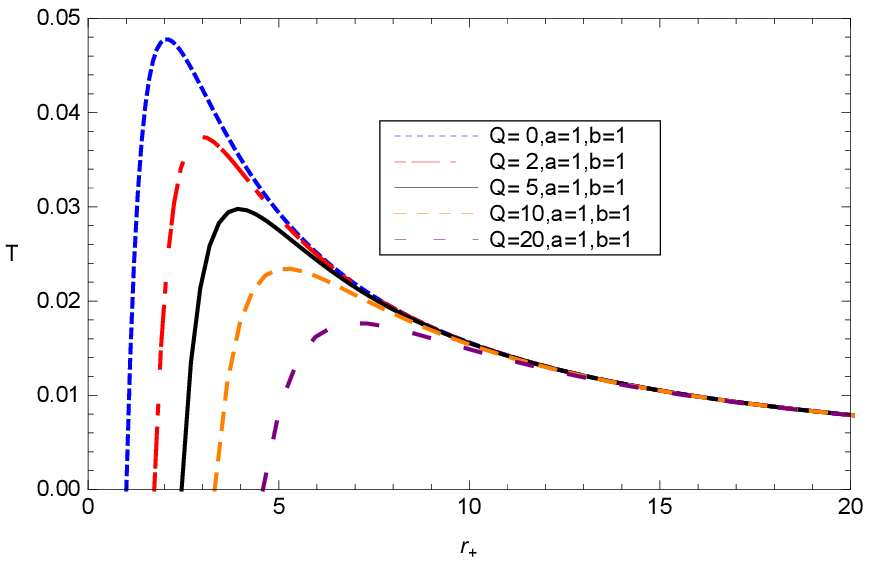}\includegraphics[width=70 mm]{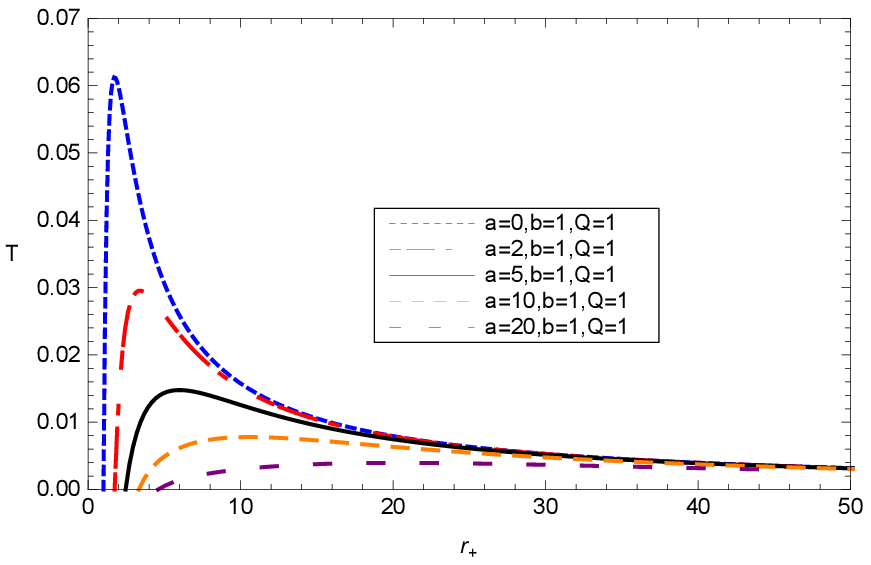}
 \end{array}$
 \end{center}
\caption{The relation between Hawking temperature and horizon $r_+$}
 \label{5}
\end{figure}

We have depicted the relation between Hawking temperature $T$ and horizon in Fig. \ref{5}, and we find the maximum value of $T$ decreases by increasing $Q$ or $a$ against the horizon position. For example we can see that a maximum temperature ($T_{Max}$) for $Q=2, a=b=1$ at about $r_{+}=3$, which means $M=6$ while $M_{c}=4$. In that case the minimum value of event horizon radius is $r_{+c}=1.732$. It means that the black hole radius, as well as the black hole mass, increases from initial values $M_{c}$ and $r_{+c}$ respectively, while the temperature also increases to $T_{Max}$. After that, by increasing the black hole size and mass, the temperature decreases as expected by Hawking's prediction. Hence, we can see unexpected behavior of temperature for smaller values of $r_{+}$. It may be due to the effects of thermal fluctuation which are important when the black hole size decreases due to the Hawking radiation  \cite{2222}.

\section{Corrected thermodynamics and Phase transition}

\begin{figure}[h!]
 \begin{center}$
 \begin{array}{cccc}
\includegraphics[width=50 mm]{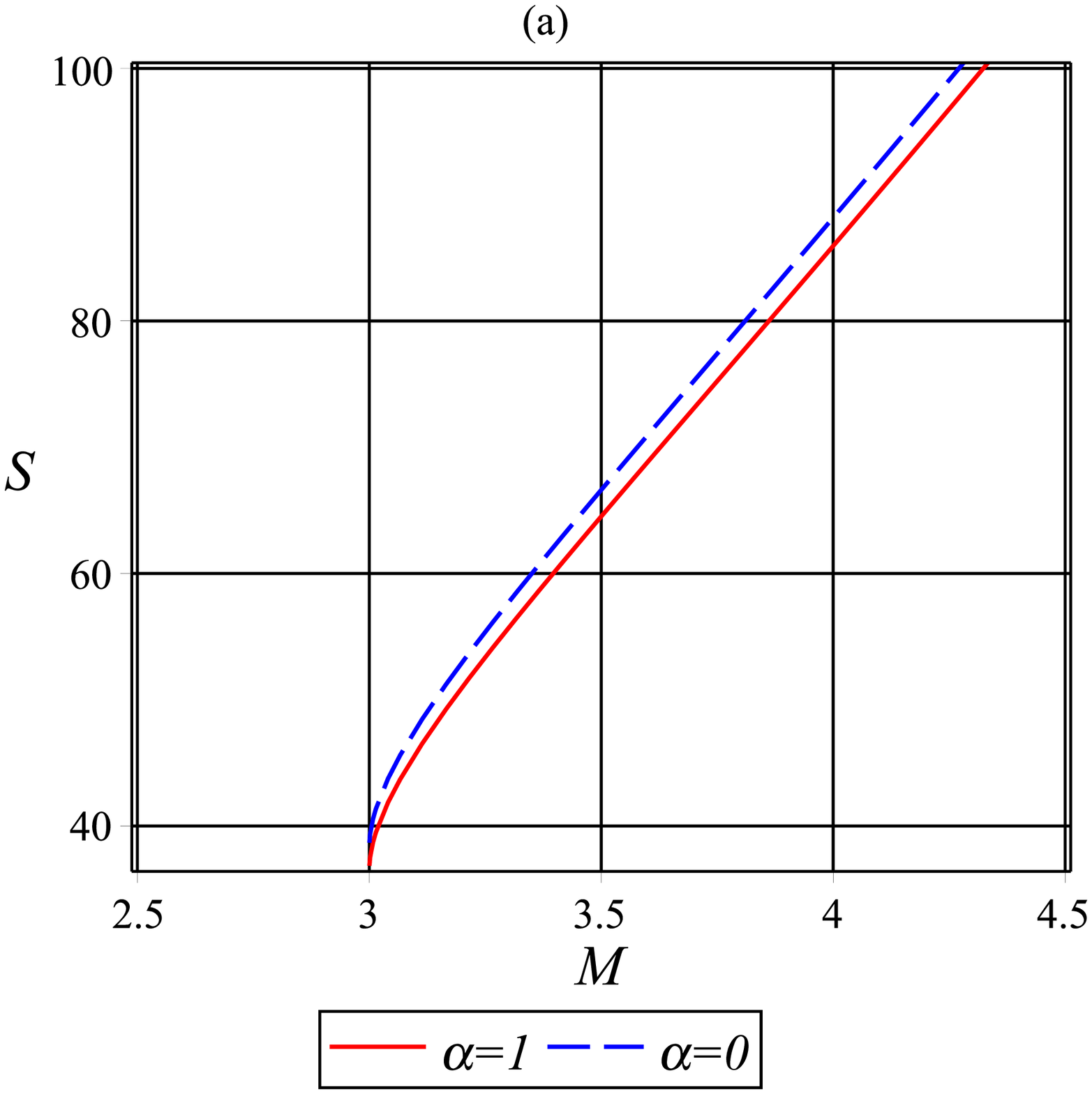}\includegraphics[width=50 mm]{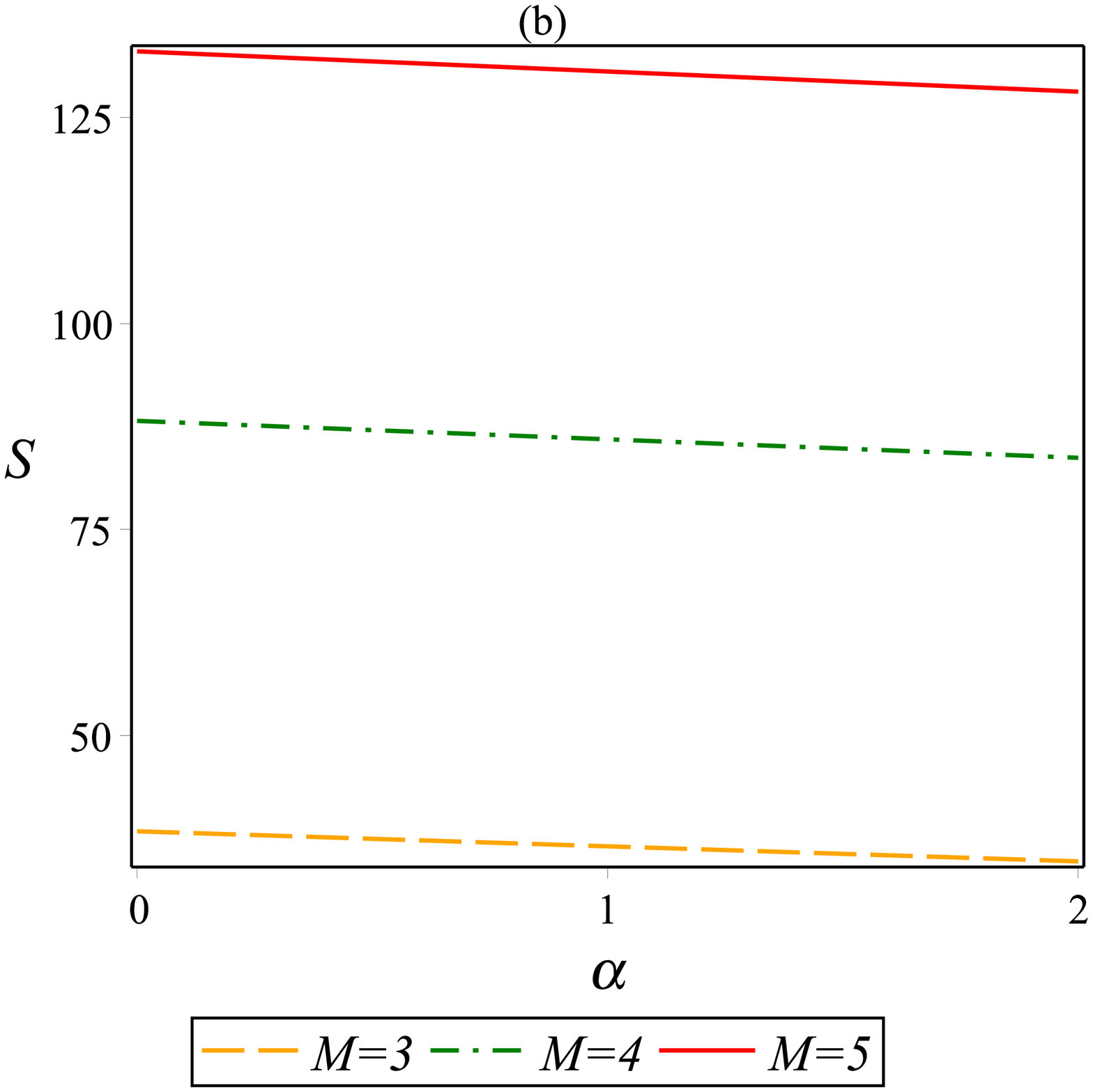}\includegraphics[width=50 mm]{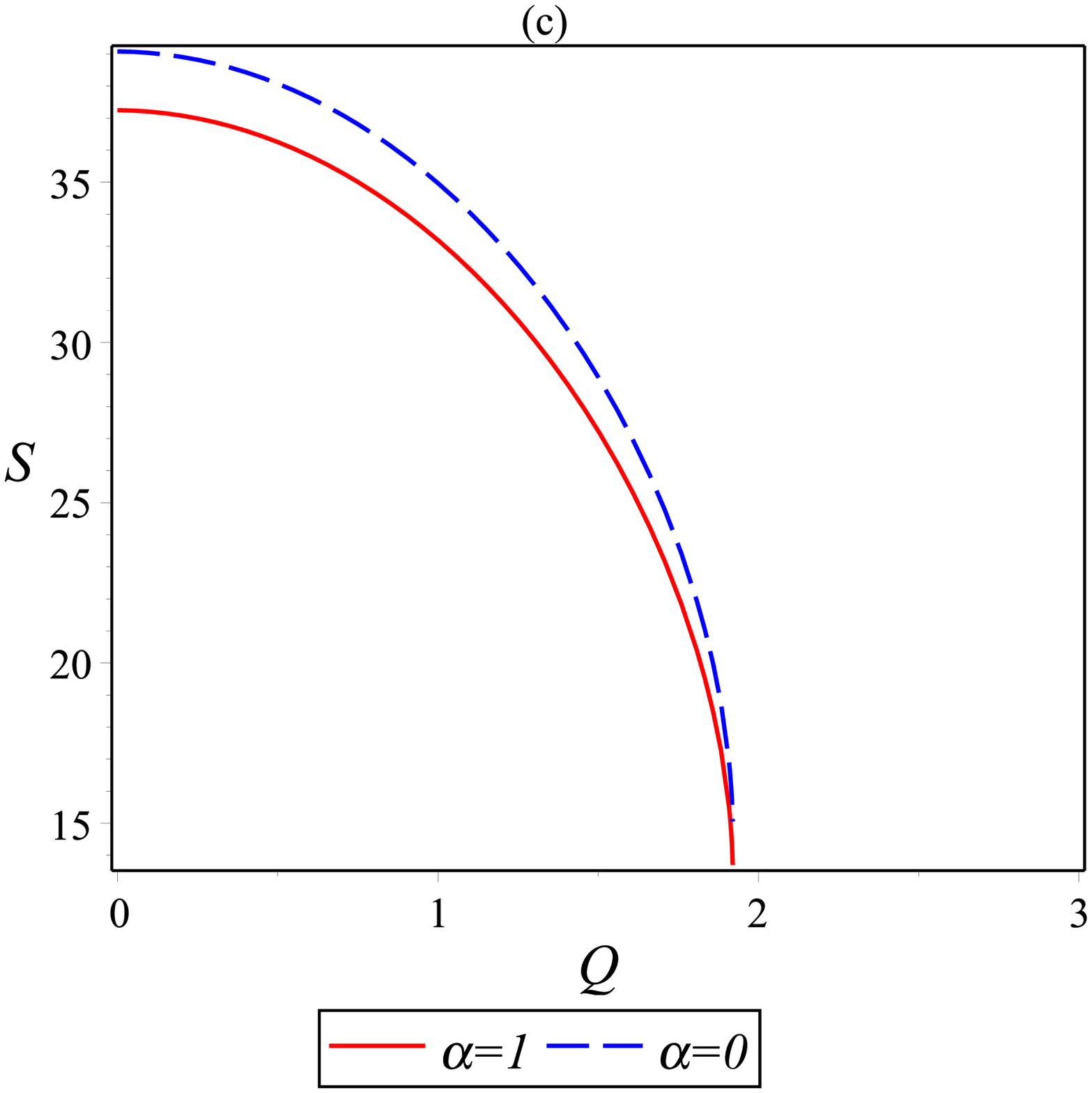}
 \end{array}$
 \end{center}
\caption{Corrected entropy in terms of (a) $M$; (b) $\alpha$ for $Q=a=b=1$; and (c) $Q$ for $a=b=0.2$, $M=2$.}
 \label{2}
\end{figure}

Due to the thermal fluctuations of space-time, the black hole entropy is modified by the logarithmic term as a first order correction \cite{sen1, sen2, M},
\begin{equation}
S=S_0-\frac{\alpha}{2}\ln S_0,
\end{equation}
where the parameter $\alpha$ is introduced by hand to track the numerical contribution of the correction term \cite{EPL}. It means that we can set $\alpha=0$ in the final result to recover classical results. Also, $S_0$ is given by
\begin{equation}\label{S}
S_0=\frac{A}{4}=\frac{\pi^2[(r_+^2+a^2)(r_+^2+b^2)+abQ]}{2r_+}.
\end{equation}
Inserting (\ref{3}) in (\ref{S}) tells us that the entropy is an increasing function of the mass parameter $M$. Hence, due to the minimum value of this parameter, there exists a minimum value for the entropy.\\
By using (\ref{S}) one can write,
\begin{equation}\label{CS}
S= \frac{\pi^2[(r_+^2+a^2)(r_+^2+b^2)+abQ]}{2r_+}- \frac{\alpha}{2}\ln \Big[  \frac{\pi^2[(r_+^2+a^2)(r_+^2+b^2)+abQ]}{2r_+} \Big].
\end{equation}
where $\alpha$ is a dimensionless correction parameter parameterizing the effect of the thermal fluctuation. In Fig. \ref{2} we observe the effects of the logarithmic correction on the black hole entropy. Notice that there is no logarithmic effect on the extremal black hole (see Fig. \ref{2} (a)). The logarithmic correction decreases the entropy (see Fig. \ref{2} (b)). Now, observe from Fig. \ref{2} (c) that the effect of the black hole charge causes the entropy to decrease. Thus the maximum entropy which corresponds to equilibrium arises when $Q=0$. Hence, we expect instability of charged MP black hole which will be discussed now by analyzing the behavior of specific heat.

We can calculate specific heat by using the following relation,
\begin{equation}
C=T\frac{dS}{dT}.
\end{equation}
In the special case of $a=b=Q=1$ we can find,
\begin{equation}
C=C_{1}C_{2},
\end{equation}
where we have defined,
\begin{eqnarray}
C_{1}&=&2\alpha\left(2M^{2}+M(r_{1}-2)-1\right)r_{2}+4\pi^{2}(3M^{2}+2)r_{1}^{2}+32M^{2}\pi^{2}(M^{2}-3M+1)-2\pi^{2}(2+r_{1}^{3})\nonumber\\
&+&8\pi^{2}r_{1}\left[6M^{3}-9M^{2}+7M+r_{1}^{3}-16-\alpha r_{2}^{3}\right]+\left(96\pi^{2}-2\alpha r_{2}^{2}\right)M,
\end{eqnarray}
\begin{equation}
C_{2}=\frac{r_{1}r_{2}}{6r_{2}^{3}+208-2M(r_{2}^{3}+112)+16M^{2}+64M^{3}-4r_{1}(r_{1}^{3}+6M^{3}-22M^{2}-40M+24)},
\end{equation}
and
\begin{eqnarray}
r_{1}&=&\sqrt{4M^{2}-8M-12},\nonumber\\
r_{2}&=&\sqrt{4M+2r_{1}-4}.
\end{eqnarray}
In Fig. \ref{3} (a) we plotted corrected specific heat in terms of $M$ and observe that there is a phase transition at a certain value $M\equiv M_{e}$. By choosing $a=b=Q=1$, we calculate $M_{e}\approx4.7$. Further, extremal black hole ($M=3$) is in the stable phase while massive black hole is in unstable phase. The effect of logarithmic correction on the phase transition is negligible, however Figs. \ref{3} (b) and (c) show variation of the specific heat with the correction parameter. For $M<M_{e}$, the specific heat is positive and logarithmic correction decreases.

\begin{figure}[h!]
 \begin{center}$
 \begin{array}{cccc}
\includegraphics[width=50 mm]{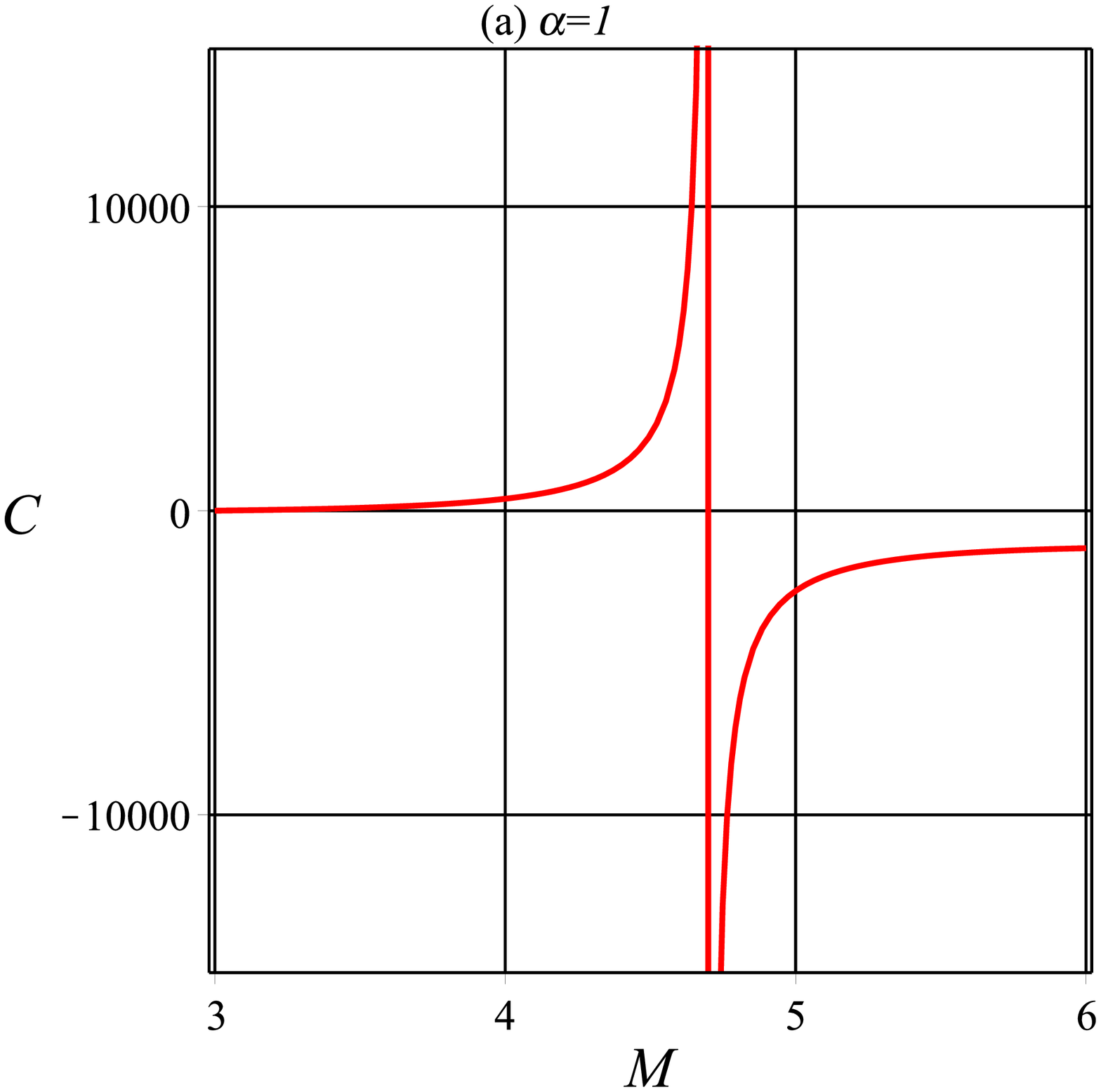}\includegraphics[width=50 mm]{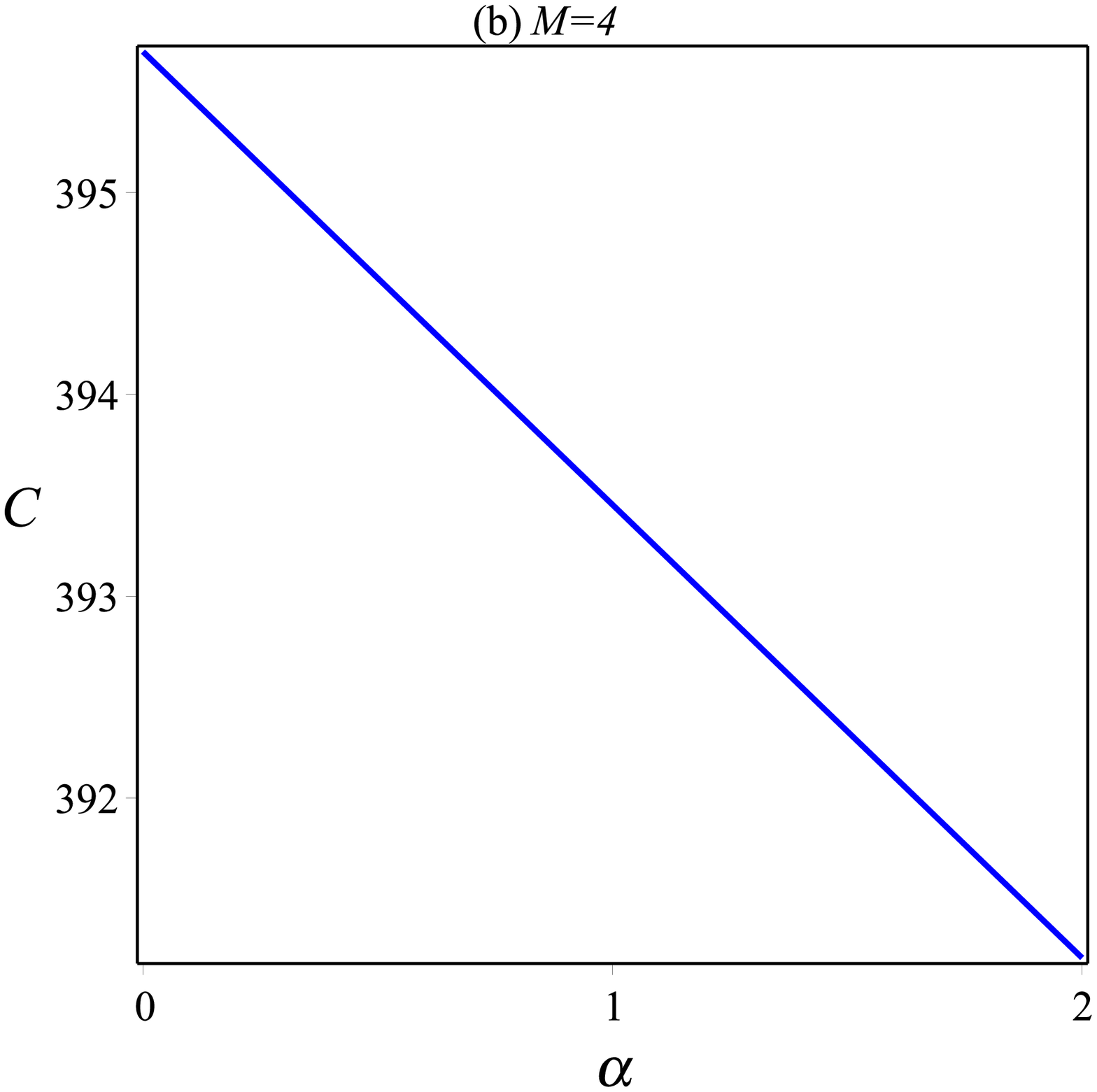}\includegraphics[width=50 mm]{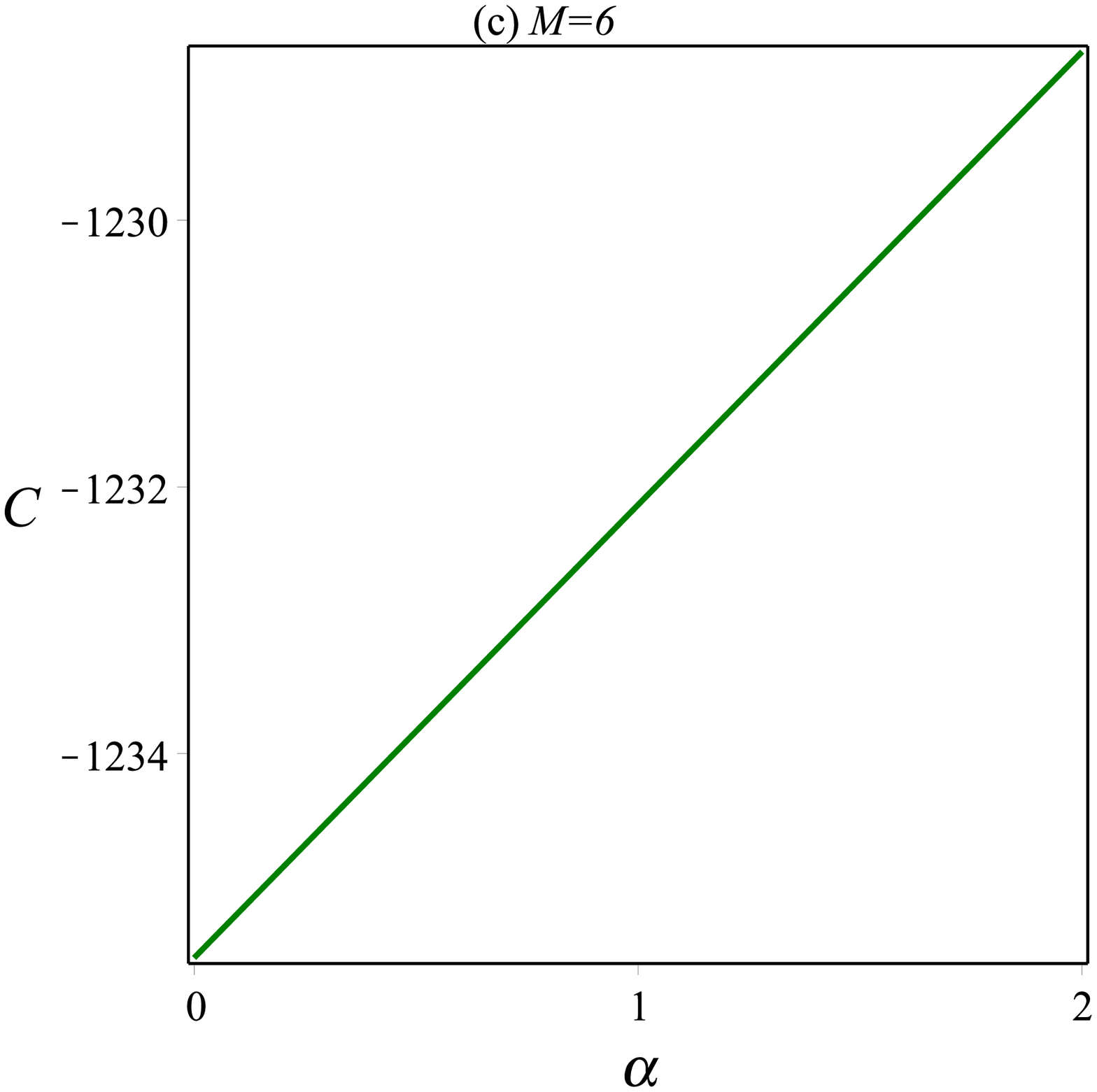}
 \end{array}$
 \end{center}
\caption{Specific heat in terms of (a) $M$, (b) and (c) $\alpha$ for $Q=a=b=1$.}
 \label{3}
\end{figure}

\begin{figure}[h!]
 \begin{center}$
 \begin{array}{cccc}
\includegraphics[width=60 mm]{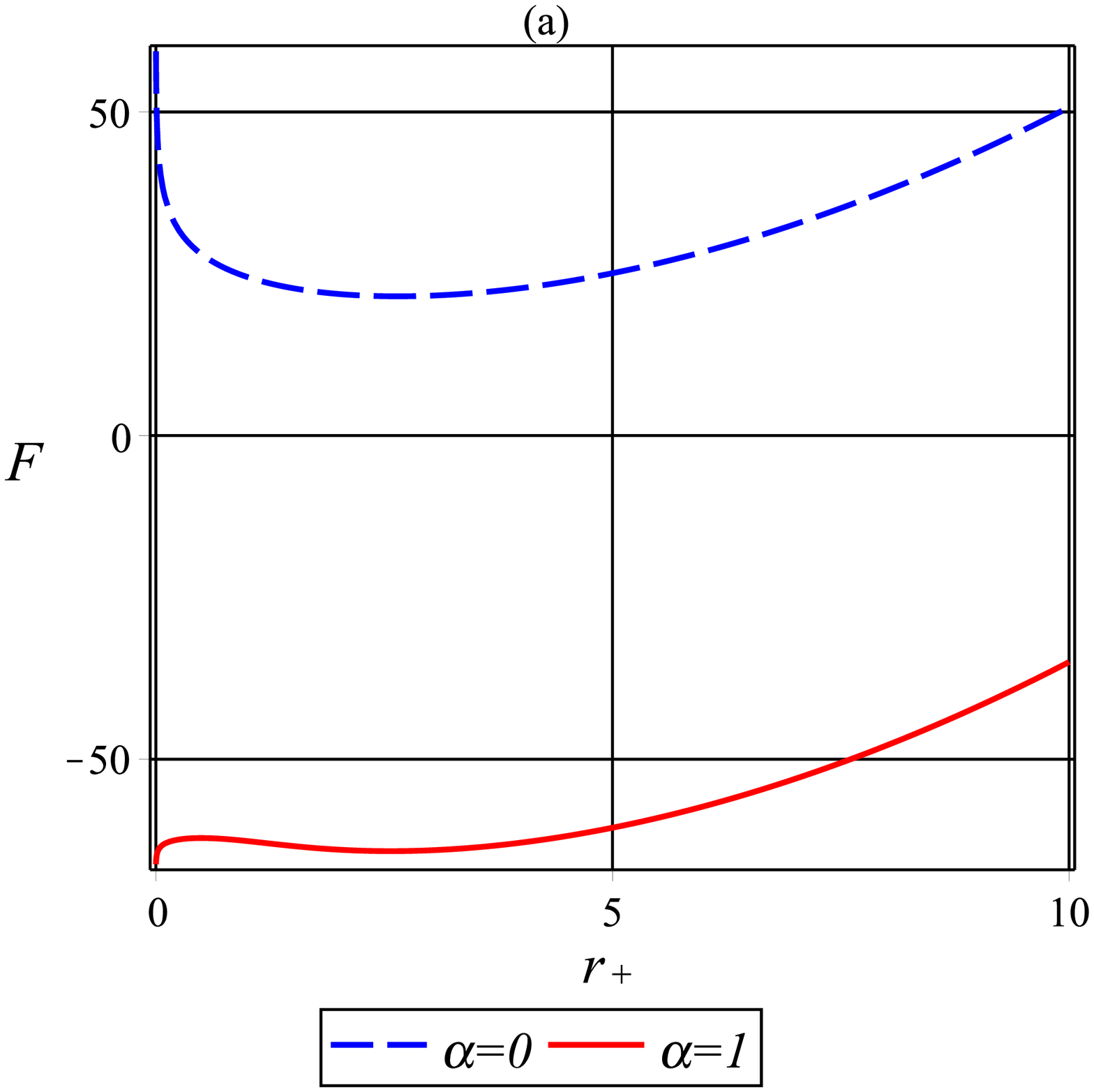}\includegraphics[width=60 mm]{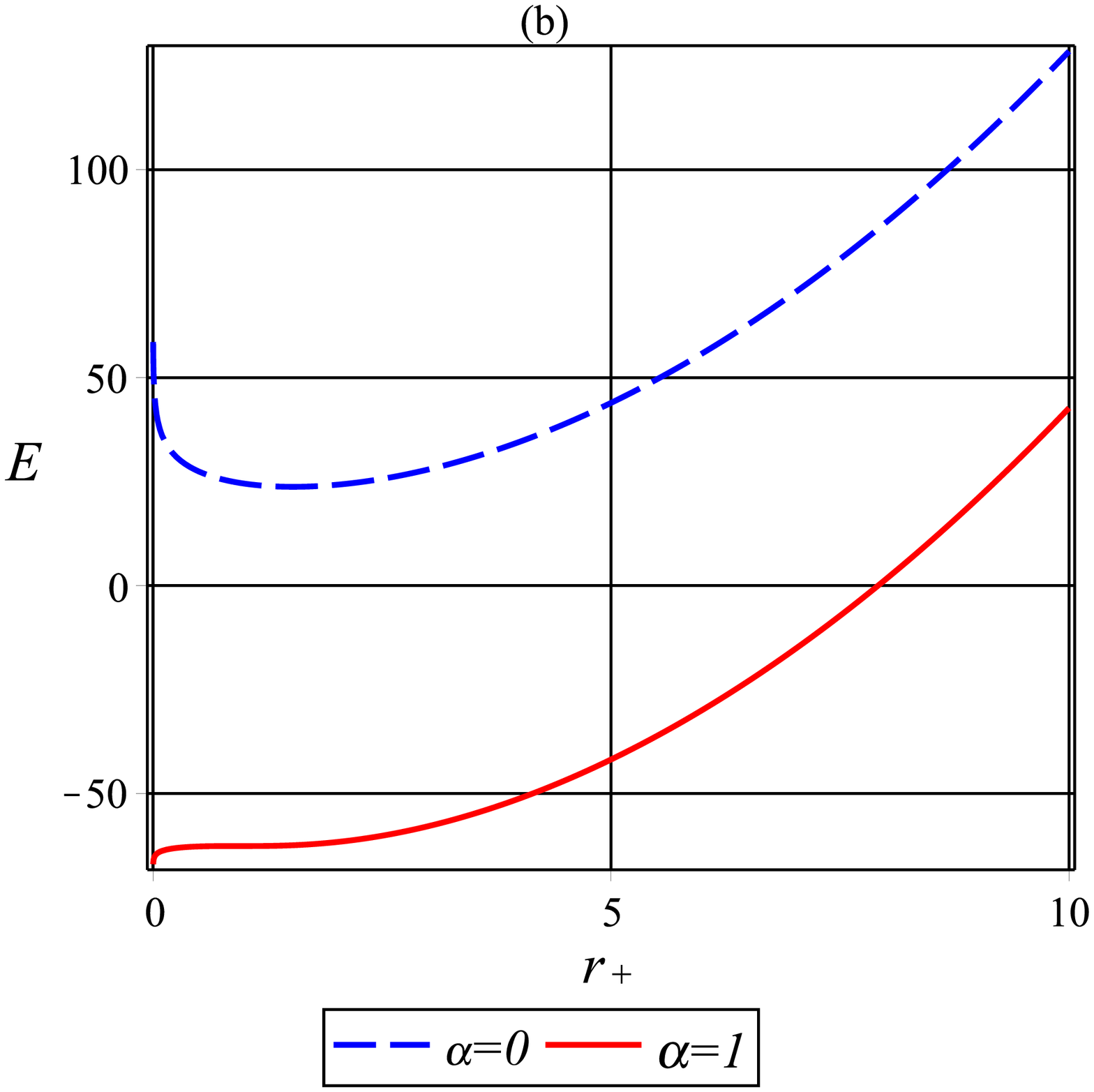}
 \end{array}$
 \end{center}
\caption{(a) Helmholtz free energy and (b) internal energy in terms of $r_{+}$ for $Q=a=b=1$ and $M=3.5$.}
 \label{4}
\end{figure}

Finally, by using the relation,
\begin{equation}
F=-\int{SdT},
\end{equation}
one can obtain Helmholtz free energy. Numerically we solve the above integral and determine important effect of logarithmic correction. In Fig. \ref{4} (a), the Helmholtz free energy in presence of logarithmic correction is negative for small $r_{+}$ and positive for the large $r_{+}$, while in the limit of $\alpha=0$ it is a completely positive quantity with a minimum. A similar behavior is observed for the internal energy which is given by $E=F+TS$ (see Fig. \ref{4} (b)).  Similar results have been reported for the Horava-Lifshitz black hole \cite{HL000}. It means that, in presence of thermal fluctuations, the black holes of five dimensional minimal supergravity behave like the black holes of Horava-Lifshitz gravity. Hence, like \cite{HL000} the second order phase transition may be possible.\\

\begin{figure}[h!]
 \begin{center}$
 \begin{array}{cccc}
\includegraphics[width=60 mm]{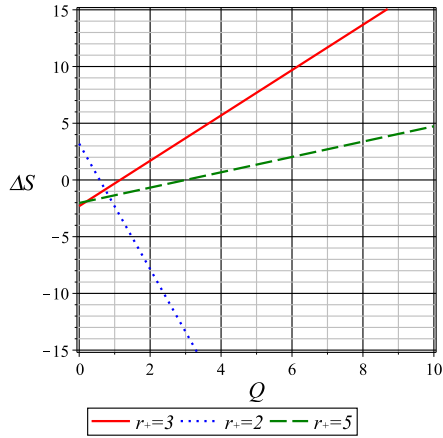}
 \end{array}$
 \end{center}
\caption{$\Delta S$ in terms of $Q$ for $a=b=0.2$, $q=1$, $\omega=0.1$  and $M=5$.}
 \label{6}
\end{figure}

The fact is that the corrected entropy (\ref{CS}) should coincide with the entropy obtained by $\Delta S=\ln{\Gamma}$ \cite{Tunneling1, Tunneling4, Tunneling5, Tunneling6} from the section III to resolve information loss paradox. Comparing it with the equation (\ref{TunnelingA}) give us the behavior of $\Delta S$ which is illustrated by Fig. \ref{6}.

We can see that increasing or decreasing of $\Delta S$ with the black hole charge is strongly dependent on the horizon radius.
It suggests that the field particle frequency should be,
\begin{equation}
\omega=\frac{\alpha\omega_{\alpha}+2\pi\omega_{Q}}{2\pi(r_{+}^{4}+(a^{2}+b^{2})r_{+}^{2}+ab(ab+Q))},
\end{equation}
where we defined
\begin{equation}
\omega_{\alpha}=(\frac{a^{2}+b^{2}}{2}+r_{+}^{2}-M)r_{+}\ln{\frac{\pi^{2}(r_{+}^{4}+(a^{2}+b^{2})r_{+}^{2}+ab(ab+Q))}{2r_{+}}},
\end{equation}
and
\begin{equation}
\omega_{Q}=\left(\frac{\sqrt{3}}{2}qQ-(aj_{a}+bj_{b})\right)r_{+}^{2}-(aj_{b}+bj_a)(ab+Q).
\end{equation}
It means that the entropy carried away by the emitted charged particle equals the original entropy of the black
hole if the above condition is satisfied, hence the Hawking radiation as tunneling presents a possibility to resolve the information loss paradox.

\section{Discussion and summary of results}
In this paper, we have constructed the formalism of Hawking radiation. We also demonstrated that various forms of field equations of quantum fields can be cast in the form of Hamilton-Jacobi equations. Thus the later equations were used to calculate the temperature of 5D minimal supergravity black hole. Using the correction terms to entropy which are motivated by thermal fluctuation, we also studied the stability of 5D minimal supergravity black hole. It was observed that this black hole is unstable because the heat capacity could be negative in some regions from Fig. \ref{3}, and the entropy of the black hole decreases with the charge and thermodynamic correction parameter. Analyzing the thermodynamics quantities shows further that in presence of quantum corrections, the black holes of five-dimensional minimal supergravity behave like the black holes of Horava-Lifshitz gravity. We also observed the phase transitions by following the change in signature of the specific heat of the black hole. We found unexpected behavior of the Hawking temperature at low mass which is due to the thermal fluctuations. When the black hole size (event horizon radius) is smaller than specific values (where $T=T_{Max}$) the black hole temperature is proportional to the black hole mass while for the larger values of $M$ and $r_{+}$ the black hole temperature is inversely proportional to the black hole mass as expected. Hence, we found quantum effects on the Hawking temperature. We compared entropy obtained by Hawking tunneling effect and logarithmic corrected entropy to obtain particle frequency. It shows the special conditions where both entropies are the same and nothing is lost in the black hole. Therefore, it may solve the information loss paradox. The logarithmic correction to the entropy of rotating and charged
BTZ black hole recently studied by Ref. \cite{SudJHAP}. For future work, it is interesting to study such an effect for the BTZ black hole in massive gravity \cite{end}. Also, in Ref. \cite{cho}, there is a non-vanishing cosmological constant $\Lambda$ which yields to the following corrected entropy
\begin{equation}\label{CS-Lambda}
S= \frac{\pi^2[(r_+^2+a^2)(r_+^2+b^2)+abQ]}{2(1-a^{2}\Lambda^{2})(1-b^{2}\Lambda^{2})r_+}- \frac{\alpha}{2}\ln \Big[  \frac{\pi^2[(r_+^2+a^2)(r_+^2+b^2)+abQ]}{2(1-a^{2}\Lambda^{2})(1-b^{2}\Lambda^{2})r_+} \Big].
\end{equation}
In that case, the black hole pressure is proportional to the cosmological constant and one can investigate whether it is equal to the thermodynamics pressure or not. Hence, it may be interesting to study $P-V$ criticality in presence of the cosmological constant. So, the black hole thermodynamics will be modified due to the cosmological constant which may be investigated as future work. Finally, it will be interesting to study non-perturbative correction \cite{exp1,exp2} on the Hawking tunneling radiation of the charged rotating black hole in five-dimensional minimal supergravity theory.

\section*{Acknowledgments}
I would like to thank Mubasher Jamil, Kai Lin, and Ge-Rui Chen for useful discussions.


\begin{thebibliography}{9}

\bibitem{mp} R. C. Myers, M.J. Perry, Annals Phys. 172, 304 (1986)

\bibitem{mp1} R. C. Myers, arXiv:1111.1903


\bibitem{111} J. An, J. Shan, H. Zhang, S. Zhao, Phys. Rev. D 97, 104007 (2018)

\bibitem{222} K. Murata, J. Soda, Prog. Theor. Phys. 120, 561 (2008)

\bibitem{333} J. An, S. Gao, [arXiv:1708.09576 [gr-qc]]

\bibitem{mp2}
P.-J. Mao, R. Li, L.-Y. Jia, J.-R. Ren, Eur. Phys. J. C 71, 1527 (2011)

\bibitem{mp3}
Z.M. Zheng, Class. Quantum Gravity 26, 135003 (2009)

\bibitem{mp4}
A. Pourdarvish and  B. Pourhassan, Int. J. Theor. Phys. 53, 136 (2014)
\bibitem{G1}
C.A.R. Herdeiro, Class. Quantum Gravity 20, 4891 (2003)
\bibitem{G2}
A. Pourdarvish, B. Pourhassan, M. Mirebrahimi, Int. J. Theor. Phys. 53, 3101 (2014)
\bibitem{G3}
A. Pourdarvish, J. Sadeghi, H. Farahani, and B.Pourhassan, Int. J. Theor. Phys. 52, 3560 (2013)
\bibitem{G4}
B. Pourhassan, K. Kokabi, Z. Sabery,  Annals of Physics 399, 181 (2018)
\bibitem{2}
T. R. Govindarajan, R. K. Kaul and V. Suneeta, Class. Quantum Grav. 18, 2877 (2001)
\bibitem{3}
J. Sadeghi, B. Pourhassan, F. Rahimi, Can. J. Phys. 92, 1638 (2014)

\bibitem{Rang} B. Pourhassan, K. Kokabi, S. Rangyan, Gen. Relativ. Gravit. 49, 144 (2017)


\bibitem{MPlog}
H. Saadat, A. Pourdarvish, Int. J. Theor. Phys. 53, 3014 (2014)


\bibitem{cho} Z. W. Chong, M. Cvetic, H. Lu, C. N. Pope, Phys. Rev. Lett. 95, 161301 (2005)



%%%%%%%%%%%%%%%%%%%%%%%%%%%%%
\bibitem{haw1} S. W. Hawking, Commun. Math. Phys. 43, 199 (1975)
\bibitem{haw2} S. W. Hawking, Nature 248, 30 (1989)
\bibitem{vis} M. Visser, JHEP 1507, 009 (2015)

\bibitem{Muk} V. Mukhanov, S. Winitzki, Introduction to Quantum  Effects in Gravity, Cambridge University Press (2010)

\bibitem{string} A. Strominger, C. Vafa, Phys. Lett. B 379, 99 (1996)

\bibitem{pady} K. Srinivasan and T. Padmanabhan, Phys. Rev. D 60, 24007 (1999)

\bibitem{mann} R. Kerner and R. Mann, Class. Quantum Grav. 25, 095014 (2008)

\bibitem{LK} K.Lin and S.Z.Yang, Chin.Phys.B 20, 110403 (2011)
\bibitem{lin1} K. Lin and S.Z. Yang, Phys.Rev.D 79, 064035 (2009)
\bibitem{lin2} K. Lin and S.Z. Yang, Phys. Lett. B 674, 127 (2009)

\bibitem{1512.04181}
A. K. Sinha, P. Majumdar, Modern Physics Letters A 32, 1750208 (2017)


\bibitem{0002040}
R. K. Kaul, P. Majumdar, Phys. Rev. Lett. 84, 5255 (2000)
%%%%%%%%%%%%%%%%%%%%%%%%%%%%%


\bibitem{86} J. Sadeghi, B. Pourhassan, M. Rostami, Phys. Rev. D 94, 064006 (2016)

\bibitem{111111} S. Upadhyay, B. Pourhassan, Prog. Theor. Exp. Phys. 2019, 013B03 (2019)



\bibitem{jcap} A.N. Aliev, JCAP 11, 029 (2014)  [arXiv:1408.4269]

\bibitem{B1}
J. Grover, J. B. Gutowski, C. A. R. Herdeiro, and W. Sabra, AIP Conference Proceedings 1122, 129 (2009)
\bibitem{B2}
Jerome P. Gauntlett, Jan B. Gutowski, Phys. Rev. D68, 105009 (2003)
%%%%%%%%%%%%%%%%%%%%%%%%%%%%%%%%
\bibitem{Tunneling1} M. Parikh and F. Wilczek, Phys. Rev. Lett. 85,5042 (2000)
\bibitem{Tunneling2} J. Zhang and Z.Zhao, Phys. Lett. B 638,110 (2006)
\bibitem{Tunneling3} R.Kerner and R.B.Mann, Class. Quan. Grav. 25,095014 (2008)

\bibitem{HJ1} K. Srinivasan and T.Padmanabhan, Phys.Rev.D 60,024007 (2000)
\bibitem{HJ2} S. Shankaranarayanan T. Padmanabhan and K.Srinivasan, Class. Quantum. Grav 19, 2671 (2002)

\bibitem{lin3} K. Lin and S.Z. Yang, Phys. Lett. B 680, 506 (2009)
\bibitem{lin4} K. Lin and S.Z. Yang, Int. J. Theor. Phys. 53, 1710 (2014)

\bibitem{m1}
R. Kerner, R.B. Mann, Class. Quant. Grav. 25, 095014 (2008)
\bibitem{m2}
R. Li, J-R. Ren, S-W. Wei, Class. Quant. Grav. 25, 125016 (2008)
\bibitem{2222}
B. Pourhassan, M. Faizal, Nuclear Physics B 913, 834 (2016)

\bibitem{sen1}
A. Sen, JHEP 04, 156 (2013)

\bibitem{sen2}
A. Sen, Gen. Rel. Grav. 44, 1947 (2012)

\bibitem{M} S. Das, P. Majumdar, R.K. Bhaduri, Class. Quant. Grav. 19, 2355 (2002)

\bibitem{EPL}
B. Pourhassan, M. Faizal, EPL 111, 40006 (2015)


\bibitem{HL000}B. Pourhassan, S. Upadhyay, H. Saadat, H. Farahani, Nuclear Physics B 928, 415 (2018)

\bibitem{Tunneling4}
Z. Niu and W. Liu, Chin. J. Phys. 46, 528 (2008)
\bibitem{Tunneling5}
L. Hui-Ling, and Y. Shu-Zheng, Commun. Theor. Phys . 51, 190 ( 2009)
\bibitem{Tunneling6}
H. Tang-Mei, and Z. Jing-Yi, Commun. Theor. Phys. 52, 619 (2009)

\bibitem{SudJHAP}
S. Upadhyay, N. Islam, P. Ganai, Journal of Holography Applications in Physics 2, 25 (2022). doi: 10.22128/jhap.2021.454.1004
\bibitem{end} B. Pourhassan, M. Faizal, Z. Zaz, and  A. Bhat,Physics Letters B 773, 325 (2017)

\bibitem{exp1}
A. Chatterjee and A. Ghosh, Phys. Rev. Lett. 125, 041302 (2020)

\bibitem{exp2}
B. Pourhassan, S. S. Wani, S. Soroushfar and M. Faizal, JHEP10, 027 (2021)

%\bibitem{proca} I. S. Landea and F. Garcia, Phys.Rev.D 94, 104006 (2016)
%\bibitem{spin2A} J.-W.Chen et al, Phys. Rev. D 81, 106008 (2010)
%\bibitem{spin2B} K. Lin et al, Mod. Phys. Lett. A 26, 1850147 (2018)

%\bibitem{GRLK} S.Z. Yang and K. Lin, Science in China, G40, 507 (2010) (in Chinese)

%\bibitem{spinN} H. Erbin and V. Lahoche, Phys. Rev. D 98, 104001 (2018)
%\bibitem{Rarita} W. Rarita and J.Schwinger, Phys. Rev. 60, 61 (1941)

\end{thebibliography}
\end{document}